\documentclass[a4paper,11pt]{article}
\usepackage{epsfig}
\newcommand{\be}{\begin{equation}}
\newcommand{\en}{\end{equation}}
\newcommand{\bea}{\begin{eqnarray}}
\newcommand{\ena}{\end{eqnarray}}
\newcommand{\lbl}[1]{\label{eq:#1}}
\newcommand{\rf}[1]{(\ref{eq:#1})}
\newcommand{\um}{u_\mu}
\newcommand{\un}{u_\nu}
\newcommand{\ua}{u_\alpha}
\newcommand{\ub}{u_\beta}

\newcommand{\cm}{\chi_-}

\newcommand{\qm}{Q_-}
\newcommand{\qp}{Q_+}
\newcommand{\fmmn}{f_{-\mu\nu}}

\newcommand{\fmgn}{f_{-\gamma\nu}}

\newcommand{\fmgb}{f_{-\gamma\beta}}

\newcommand{\fpmn}{f_{+\mu\nu}}
\newcommand{\fpna}{f_{+\nu\alpha}}
\newcommand{\fpab}{f_{+\alpha\beta}}
\newcommand{\fpgm}{f^\gamma_{+\mu}}

\newcommand{\fpga}{f^\gamma_{+\alpha}}

\newcommand{\hgn}{h_{\gamma\nu}}

\newcommand{\hgb}{h_{\gamma\beta}}

\newcommand{\trace}[1]{\langle #1 \rangle}
\newcommand{\com}[1]{ [ #1 ]}
\newcommand{\la}{\langle}
\newcommand{\ra}{\rangle}
\newcommand{\eeps}{\epsilon^{\mu\nu\alpha\beta} }
\newcommand{\ceps}{\kappa\,   \epsilon^{\mu\nu\alpha\beta} }
\newcommand{\deps}{\kappa\,\hat\epsilon_{\mu\nu\alpha\beta} }
\newcommand{\pitl}{\nabla_\mu\pi^3}
\newcommand{\pitlnu}{\nabla_\nu\pi^3}
\newcommand{\piel}{\nabla_\mu\pi^8}
\newcommand{\pielnu}{\nabla_\nu\pi^8}
\newcommand{\pith}{\nabla^\mu\pi^3}
\newcommand{\pieh}{\nabla^\mu\pi^8}
\newcommand{\mpid}{m^2_\pi}
\newcommand{\fpid}{F^2_0}
\newcommand{\fpiq}{F^4_0}
\newcommand{\mkd}{m^2_K}
\newcommand{\undemi}{ {1\over2}}

\newcommand{\ow}[1]{\hat o^W_{#1} }
\newcommand{\const}{{\displaystyle\kappa\over\displaystyle  F^2}}
\begin{document}

\begin{titlepage}
\begin{flushright}
IISc-CTS-03/02\\
IPNO/DR-02-013/\\
\end{flushright}

\begin{center}
{\bf ELECTROMAGNETIC CORRECTIONS IN THE ANOMALY SECTOR}\\[1.5cm]

{\large B. Ananthanarayan$^a$  and B. Moussallam$^b$}\\[1cm]

{\sl$^a$\  Centre for Theoretical Studies}\\
{\sl Indian Institute of Science}\\
{\sl Bangalore 560 012, India}

{\sl$^b$\ Groupe de Physique Th\'eorique, IPN }\\
{\sl Universit\'e Paris-Sud}\\
{\sl F-91406 Orsay C\'edex, France}
\end{center}
\vfill

\begin{abstract}
Chiral perturbation theory in the anomaly sector for $N_f=2$ is extended 
to include dynamical photons, 
thereby allowing a complete treatment of isospin breaking.
A minimal set of independent chiral lagrangian terms is determined
and the divergence structure is worked out. 
There are contributions from irreducible and also from reducible one-loop
graphs, a feature of ChPT at order larger than four. 
The generating functional is non-anomalous at order $e^2p^4$, but
not necessarily at higher order in $e^2$. 
Practical applications to $\gamma\pi\to\pi\pi$ 
and to the $\pi^0\to2\gamma$ amplitudes are considered. In the latter case,
a complete discussion of the corrections beyond current algebra is presented
including  quark mass as well as electromagnetic effects.  
\end{abstract}
\vfill
\end{titlepage}

\setcounter{page}{1}
\setcounter{equation}{0}
\section{Introduction}

The precise determination of the masses of the two lightest quarks $m_u$, 
$m_d$ in the standard model 
using chiral perturbation theory (ChPT)
requires one to consider small  isospin breaking phenomena
(see e.g.\cite{leutw} for a review on the present knowledge of the
light quark masses). In this case, 
it is necessary to account for  radiative corrections  which
induce isospin breaking effects of comparable size to that induced 
by the quark mass difference $m_d-m_u$. In practice, this amounts to including
the photon as a dynamical field, along with those of the quasi-goldstone
bosons into the chiral Lagrangian. This extension of the conventional
ChPT framework\cite{W,gl84,gl85}, 
in the ordinary sector of ChPT, is due to Urech\cite{urech}. Recently,
inclusion of dynamical photons was discussed in the sector of the weak
non-leptonic interactions\cite{ecker}. In the present paper we will discuss
the odd intrinsic-parity sector of ChPT. 

One motivation for this study is the celebrated $\pi^0\to2\gamma$ process. In 
the chiral limit the amplitude is non-vanishing and exactly predicted from
the ABJ anomaly\cite{adler,belljackiw}. Away from this limit, the 
amplitude receives small corrections from $m_u$, $m_d$ which are dominated
by the isospin breaking difference $m_d-m_u$ as well as 
electromagnetic corrections. Comparison with high-precision
measurements of the $\pi^0$ lifetime, such as the PrimEx 
experiment\cite{primex} at JLAB (which plans to achieve a level of precision
of less than a percent) could provide a new insight into $m_d-m_u$ and,
more generally, furnish a new test of ChPT. 
An explicit calculation of radiative corrections at one-loop was recently
performed in ref.\cite{akt} for the anomalous amplitude
$\gamma\pi^+\to  \pi^+\pi^0$. Our work should complement and provide a 
check of this calculation. Analogous calculations might also be needed for
other precisely determined pionic observables, for instance, the 
$\beta$-decay vector form factor. As far as only pions are concerned, 
the $SU(2)$ chiral expansion is the most general one. 

The plan of the paper is as follows. For kinematical reasons, in the odd
intrinsic parity sector the leading chiral order is $p^4$. At this order, 
the effective action in this sector is anomalous, i.e. it is not 
invariant under
chiral transformations and its variation under infinitesimal 
axial transformations must reproduce the Bardeen 
expression for the anomaly\cite{bardeen}. The explicit form of the anomalous
effective action, the Wess-Zumino-Witten  action, is known for
an arbitrary number of chiral flavours $N_f$\cite{wz,hd,witten,morewz}.
Recently, a  simpler form was derived for the particular case
$N_f=2$\cite{kaiser} which is used in the present paper.
Using the 
spurion technique, we have first classified the minimal set of chiral 
Lagrangian terms of order $e^2p^4$ in the case of $SU(2)$ ChPT. We find
that there are eight independent terms and corresponding new chiral 
coupling-constants $k^W_i$. We next consider the one-loop 
contributions to the generating functional of order $e^2p^4$ in this 
sector.  We observe that while the generating functional is
non-anomalous at order $e^2p^4$ this is no longer true at higher order. 
We  compute its divergence structure and the corresponding 
renormalization  of the coupling-constants $k^W_i$ and these results
are applied to the $\gamma\pi^+\to  \pi^+\pi^0$ amplitude. 
Next, we discuss the 
electromagnetic corrections to the $\pi^0$ decay amplitude. In this case, it
turns out that the couplings $k^W_i$ make no contributions, the only 
contributions are generated by one-particle reducible graphs. We derive
the expression for the amplitude incorporating the complete set of corrections
linear in the quark masses and quadratic in the electric charge. We finally
discuss how to exploit this expression by combining resonance saturation
estimates and the $SU(3)$ chiral expansion.

\section{Chiral Lagrangian at order $e^2p^4$}

From the Adler-Bardeen theorem\cite{bardeen1} we expect the chiral Lagrangian
in the odd-intrinsic parity sector at order $e^2 p^4$ to be non-anomalous
\footnote{The general analysis of ref.\cite{bardeen1} of the QED 1PI
diagrams shows that all anomalous contributions are generated
from diagrams having the one-loop triangle diagram as a sub-graph. Compared
to the leading order triangle diagram any contribution of this type is
suppressed by a factor $e^4$ . }. In order to construct the various terms
in the chiral Lagrangian we may therefore employ the same type of chiral
building blocks as in the ordinary sector.

\subsection{Building blocks}

Virtual photons induce new chiral Lagrangian terms which include factors
of the electric charge matrix $Q$, which for $N_f=2$ is
\be
Q=e\left(
\begin{array}{cc}
{2\over3} & 0 \\
0         &-{1\over3}\\
\end{array}\right)\ .
\en
One can think of these terms as encoding the high-energy part of the photon
loops.  In order to perform a consistent chiral expansion one
must define a chiral counting rule for $Q$. We will follow the choice
proposed in ref.\cite{urech}
\be
O(e)\equiv O(p)\ .
\en
Next, in order to classify the number of independent chiral Lagrangian
terms at a given chiral order, one replaces the charge matrix $Q$ by two
charge spurions endowed with the transformation properties
\be
Q_L\to g_L Q_L g_L^\dagger,\quad Q_R\to g_R Q_R g_R^\dagger\ .
\en
Here, we will be interested in classifying the terms of order $e^2p^4$
in the odd intrinsic-parity sector. A convenient set of building-blocks
for the classification of higher order chiral Lagrangian terms\cite{bce1}
are those associated with a non-linear representation of the chiral 
group\cite{cwz}. We will follow the  notation of ref.\cite{bce1}
(see also \cite{egpr}) which is summarized below. Given the usual unitary 
matrix $U$  encoding 
the pseudo-scalar fields and which transforms as $U\to g_R^\dagger U g_L$ one
may generate a non-linear representation by setting
\be
U=u^2,\quad u\to g_R^\dagger u h=h^\dagger u g_L\ .
\en
A building-block $X$, then, transforms in the following way under a chiral
transformation,
\be
X\to h X h^\dagger\ .
\en
A connection $\Gamma_\mu$ may be introduced such that the covariant derivative
\be
\nabla_\mu X=\partial_\mu X +[\Gamma_\mu,X],\ 
{\rm with} \ 
\Gamma_\mu={1\over2}\left[u^\dagger (\partial_\mu-i r_\mu)u 
+u         (\partial_\mu-i l_\mu)u^\dagger\right]\ 
\en
is also a building block. Another building-block is the field-strength
tensor $\Gamma_{\mu\nu}$,
generated from the
commutator of two covariant derivatives, 
\be\lbl{commut}
[\nabla_\mu, \nabla_\nu] X=[\Gamma_{\mu\nu},X]\ .
\en
It satisfies the Bianchi Identity, which is important in the odd 
intrinsic-parity sector\cite{aa},
\be\lbl{bianchi}
\nabla_\alpha\Gamma_{\mu\nu}+\nabla_\mu\Gamma_{\nu\alpha}+
\nabla_\nu\Gamma_{\alpha\mu}=0\ .
\en
The following standard building-blocks which will appear in our construction
\be
u_\mu=u^\dagger_\mu=i u^\dagger D_\mu U u^\dagger
=i\left(u^\dagger (\partial_\mu-i r_\mu)u 
-u(\partial_\mu-i l_\mu)u^\dagger\right)
\en
and 
\be
f_\pm^{\mu\nu}= u f^{\mu\nu}_L u^\dagger \pm u^\dagger f^{\mu\nu}_R u\ .
\en
In our construction all the Lorentz indices which appear are contracted
with the antisymmetric $\epsilon$ tensor. It is therefore not necessary
to consider more than one derivative acting on one building-block because
of relation \rf{commut}. Also $\nabla_\mu u_\nu$ need not
be considered separately because of the relation
\be\lbl{fmin}
\nabla_\mu u_\nu-\nabla_\nu u_\mu=-f_{-\mu\nu}\ . 
\en
We also recall the following relation
\be
\Gamma_{\mu\nu}={1\over4}[u_\mu,u_\nu]-{i\over2}f_{+\mu\nu}
\en
which indicates that $\Gamma_{\mu\nu}$ need not be independently considered.

From the charge spurions, one can form the building blocks
\be
Q_\pm= u Q_L u^\dagger \pm u^\dagger Q_R u
\en
and one can also consider their covariant derivatives. We will, instead,
use the following building blocks
\be
Q^\mu_\pm= u c^\mu_L Q_L u^\dagger \pm u^\dagger c^\mu_R Q_R u
\en
in which the derivatives acting on the charge spurions are
defined as\cite{urech}
\be
c^\mu_L Q_L=\partial^\mu Q_L -i[l_\mu,Q_L],\quad
c^\mu_R Q_R=\partial^\mu Q_R -i[r_\mu,Q_R],\quad\ .
\en
They are related to  covariant derivatives by the following relation
\be\lbl{qmu}
Q^\mu_\pm=\nabla^\mu Q_\pm+{i\over2} [u^\mu,Q_\mp]\ .
\en
This completes the list of the chiral building-blocks needed in our
construction.
Their transformation properties
under parity and charge conjugation are collected in table 1 .
\begin{table}[hbt]
\begin{center}
\begin{tabular}{|c|c c c c |}\hline
 \ & $u_\mu$  & $f_{\pm\mu\nu}$      & $Q_{\pm}$     & $Q_{\pm\mu}$\\ \hline
 P & $-u^\mu$ & $\pm f^{\mu\nu}_\pm$ & $\pm Q_{\pm}$ & $\pm Q^\mu_{\pm}$
\\ \hline
 C &$u^T_\mu$ & $\mp f^T_{\pm\mu\nu}$& $\mp Q^T_{\pm}$ & $\mp Q^T_{\pm\mu}$
\\ \hline
\end{tabular}
\caption{ Transformation properties of the chiral building-blocks
under parity (P) and charge conjugation (C).}
\end{center}
\label{Table 1}
\end{table}

\subsection{Classification of a minimal set of independent terms}

Our purpose is to construct a complete set of chiral Lagrangian terms,
which are chirally invariant, and of order $e^2 p^4$ in the odd 
intrinsic-parity sector. We must then have four lorentz indices 
contracted with the $\epsilon$ tensor and two insertions of the 
charge matrix $Q$. One can a priori form seven independent types
of terms of this kind
\be
\begin{array}{lll}
 1)\ \eeps Q Q u_\mu u_\nu u_\alpha u_\beta,\ 
&2)\ \eeps Q Q f_{\mu \nu} u_\alpha u_\beta,\ 
&3)\ \eeps Q Q f_{\mu \nu} f_{\alpha \beta},\ \\
 4)\ \eeps Q_\mu Q u_\nu u_\alpha u_\beta,\ 
&5)\ \eeps Q_\mu Q u_\nu f_{\alpha \beta},\ 
&6)\ \eeps Q_\mu Q_\nu u_\alpha u_\beta,\  \\
 7)\ \eeps Q_\mu Q_\nu f_{\alpha\beta}
\end{array}
\en
Here we have not specified the $\pm$ subscripts and all the possible 
independent orderings and 
trace structures are to be worked out later. Using 
eq.\rf{qmu} we see that by using integration by parts we can always express
the structures 6) and 7) in terms of the others, so only the first five 
structures are needed . 

We specialize to the case of $SU(2)$ ChPT in this section, which allows one to
discuss electromagnetic corrections for processes involving pions. 
For hermitian 2x2 
matrices the following identity relates the anti-commutator to traces
\be\lbl{ident}
\{A,B\}=A\la B\ra  +B\la A\ra +\la AB\ra -\la A\ra \la B\ra . 
\en
In connection with this identity, one notes that the product of three
matrices can be written in terms of anti-commutators,
\be\lbl{abc}
ABC={1\over2}\left( \{A,B\}C+A\{B,C\}-\{B,AC\}\right)\ .
\en
For classification purposes, it is thus not necessary to consider traces
of products of more than three matrices.
In the case of $SU(2)$ the trace of the charge matrix does not vanish but we
can restrict ourselves to $\la Q_L\ra =\la Q_R\ra $ 
(following ref.\cite{ku}) i.e.
\be
\la Q_-\ra =0\ .
\en
Also we will assume that $\la \partial_\mu Q_L\ra =\la \partial_\mu Q_R\ra =0$, i.e.
\be
\la Q_{\pm\mu}\ra =0\ .
\en
Concerning the external vector and axial-vector fields we must consider the
case $\la f_{+\mu\nu}\ra \ne 0$ but we can restrict ourselves to
$\la f_{-\mu\nu}\ra =0$. 
Let us now start the classification of independent terms.

$\bullet${Terms of type (1)}

Making use of the relations \rf{ident} and \rf{abc} it is easy to see
that there is a single term of this kind,
\be
\ow{1}\equiv
i\eeps \la Q_+\ra \la Q_- u_\mu\ra \la u_\nu u_\alpha u_\beta\ra 
\en
(the factor of $i$ ensures hermiticity)

$\bullet${Terms of type (2)}

We begin by considering terms containing $Q_+$, $Q_-$, $f_{+\mu\nu}$. 
One finds that there are two independent terms of this type, which we
choose as
\bea
&&\ow{2}\equiv\eeps \la [Q_+,Q_-]u_\alpha u_\beta\ra \la f_{+\mu\nu}\ra 
\nonumber\\
&&\ow{3}\equiv\eeps \la Q_+\ra \la [Q_-,f_{+\mu\nu}]u_\alpha u_\beta\ra 
\ena
Next,
we consider the terms containing $Q_+$, $Q_+$, $f_{-\mu\nu}$, we find only
one independent term of this kind
\be
\ow{4}\equiv\eeps \la Q_+^2 u_\alpha\ra \la  f_{-\mu\nu} u_\beta\ra 
\en
Finally, we consider the terms containing $Q_-$, $Q_-$, $f_{-\mu\nu}$:
all possible terms of this kind which are C-even turn out to vanish.

$\bullet${Terms of type (3)}

We consider first the combinations $Q_+ Q_- f_{+\mu\nu} f_{+\alpha\beta}$ .
There is only one C-even double trace term
\be
\ow{5}\equiv i\eeps \la [Q_+,f_{+\mu\nu}]Q_-\ra \la f_{+\alpha\beta}\ra 
\en
and no C-even triple trace terms. Next, the combination
$Q_+ Q_- f_{-\mu\nu} f_{-\alpha\beta}$ gives rise to no contribution
because we restrict ourselves to $\la f_{-\mu\nu}\ra =0$. What remains to
be considered are the two combinations  $Q_+ Q_+ f_{+\mu\nu} f_{-\alpha\beta}$
and $Q_- Q_- f_{+\mu\nu} f_{-\alpha\beta}$. Only one invariant 
is generated from these
\be
\ow{6}\equiv i\eeps \la Q_+\ra \la [Q_+,f_{+\mu\nu}]f_{-\alpha\beta}\ra \ .
\en

$\bullet${Terms of type (4)}

Using parity invariance one needs consider the combination $Q_{+\mu} Q_+$ 
or $Q_{-\mu} Q_-$, it turns out that no C-even term of type (4) can be 
formed.

$\bullet${Terms of type (5)}

One finds four independent terms of this type
\bea
&&\ow{7}\equiv i\eeps\la [Q_{+\mu}, Q_+]u_\nu\ra \la f_{+\alpha\beta}\ra 
\nonumber\\
&&\ow{8}\equiv i\eeps\la [Q_{-\mu}, Q_-]u_\nu\ra \la f_{+\alpha\beta}\ra 
\ena
and
\bea
&&\ow{9}\equiv\eeps \la  [Q_{+\mu},u_\nu] f_{+\alpha\beta}\ra \la Q_+\ra 
\nonumber\\
&&\ow{10}\equiv\eeps\la  [Q_{-\mu},u_\nu] f_{-\alpha\beta}\ra \la Q_+\ra \ .
\ena
The last two terms need not, however, be considered because
using integration by parts they can be related to terms already considered
and to terms containing $\la Q_{+\mu}\ra $ which was assumed to vanish. 
Finally, one must worry about contact terms, i.e. invariants involving
external fields only. No such terms can be formed here which respect
$P$ and $C$ invariance. 

In conclusion, we find that there are eight independent chiral Lagrangian terms
of order $e^2 p^4$, in the odd 
intrinsic-parity sector. We may label the associated chiral coupling
constants as $k^W_i$ by analogy with the couplings $k_i$ introduced
for the $SU(2)$ chiral Lagrangian at order $e^2p^2$ \cite{ku}. The
resulting  chiral Lagrangian is summarized below,
\be\lbl{lwe2p4}
\begin{array}{lll}
{\cal L}^W_{e^2p^4} = 
\eeps\Big\{ 
  &\phantom{+}  
     k^W_1\,   i\la Q_+\ra \la Q_- u_\mu\ra \la u_\nu u_\alpha u_\beta\ra 
{}&+ k^W_2\,   \la [Q_+,Q_-]u_\alpha u_\beta\ra \la f_{+\mu\nu}\ra\\
\noalign{\vskip0.2truecm}
{}&+ k^W_3\,   \la Q_+\ra \la [Q_-,f_{+\mu\nu}]u_\alpha u_\beta\ra 
{}&+ k^W_4\,   \la Q_+^2 u_\alpha\ra \la  f_{-\mu\nu} u_\beta\ra \\
\noalign{\vskip0.2truecm}
{}&+ k^W_5\,    i\la [Q_+,f_{+\mu\nu}]Q_-\ra \la f_{+\alpha\beta}\ra 
{}&+ k^W_6\,    i\la Q_+\ra \la [Q_+,f_{+\mu\nu}]f_{-\alpha\beta}\ra \\
\noalign{\vskip0.2truecm}
{}&+ k^W_7\,    i\la [Q_{+\mu}, Q_+]u_\nu\ra \la f_{+\alpha\beta}\ra 
{}&+ k^W_{8}\, i\la [Q_{-\mu}, Q_-]u_\nu\ra \la f_{+\alpha\beta}\ra \Big\}
\end{array}
\en

\section{Divergence structure}

\begin{figure}[abt]
\epsfysize=9cm
\begin{center}
\epsffile{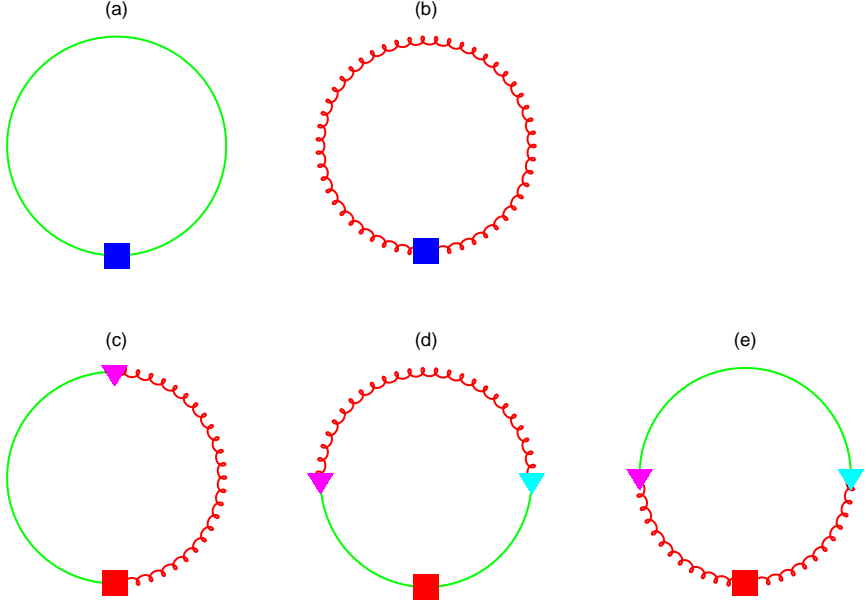}
\caption{One particle irreducible graphs
which contribute to the $O(p^6)$ and the $O(e^2p^4)$ divergences
in the anomalous sector: a triangle
denotes an $O(p^2)$ vertex a box denotes an $O(p^4)$ vertex. The solid
line represents the (full) pion propagator and the curly line the full
photon propagator.}
\label{Fig. 1}
\end{center}
\end{figure}
In order to determine the divergent parts of the generating functional
$S^W_{e^2p^4}$ we can proceed by analogy with the general discussion 
of Bijnens, Colangelo and Ecker (BCE) \cite{bce2} 
of the $O(p^6)$ divergence structure of ChPT. 
Since we concern ourselves with the odd intrinsic-parity 
sector here, only one-loop diagrams will contribute which have
one vertex from the WZW action $S^W_{p^4}$.  The one-particle irreducible
diagrams are shown in fig.1. Diagram (a) does not contain the photon
explicitly but a contribution proportional to $e^2$ can get generated
from the $\pi^+-\pi^0$ mass difference, also proportional to the 
low-energy coupling $C$. The divergent contribution from this diagram 
when $e=0$ was evaluated previously for generic $N_f$ and traceless
external vector or axial-vector 
fields \cite{wyler,bramon,issler,aa} and in the particular
case of $N_f=2$ and non-traceless external vector fields in ref.\cite{bgt}. 

As pointed out by BCE it is also necessary to consider one-particle
reducible graphs. Such graphs are shown in fig.2. The origin of these extra
divergences is that the $O(p^4)$ chiral Lagrangian cancels the $O(p^4)$
one-loop divergences only up to terms which vanish upon using the equations
of motion. The form of these  depends on the choice of independent
terms in the $O(p^4)$ chiral Lagrangian. We will next display the
expressions for the various vertices involved and then give the results
of computing the diagrams of figs.1 and 2.

\begin{figure}[abt]
\epsfysize=6cm
\begin{center}
\epsffile{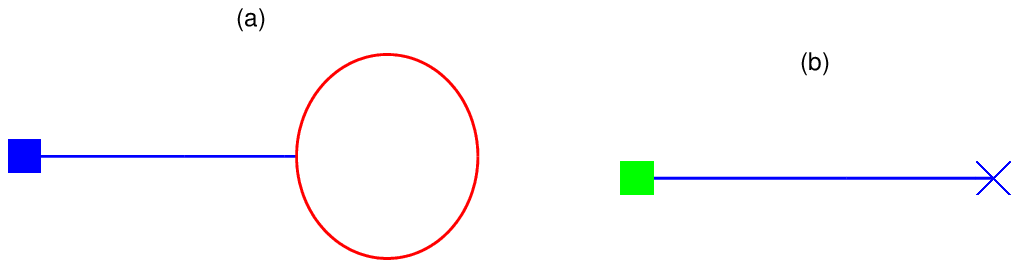}
\caption{Sample one particle reducible graphs
which can contribute to the $O(p^6)$ and the $O(e^2p^4)$ divergences.
A square denotes a vertex from the WZW Lagrangian, a cross a vertex
from the $O(p^4)$ Lagrangian. }
\label{Fig. 2}
\end{center}
\end{figure}
\subsection{$O(p^2)$ vertices}

One starts from the $O(p^2)$ chiral Lagrangian including dynamical
photon fields\cite{urech},
\bea
&{\cal L}_{p2}=&{F^2\over4}\langle D_\mu U D^\mu U^\dagger +\chi^\dagger U
+U^\dagger \chi\rangle
+C\langle Q_R U Q_L U^\dagger\rangle
\nonumber\\
&&\quad-{1\over4} A_{\mu\nu} A^{\mu\nu}-{1\over2\lambda} (\partial^\mu
A_\mu)^2
\ena
The photon field $A_\mu$ also appears in the covariant derivative
\be
D_\mu U=\partial_\mu U-ir_\mu U+iU l_\mu +iA_\mu(-Q_R U+U Q_L)\ .
\en
The fluctuations of the pion field around the classical 
configuration may be defined by setting\cite{gl85}
\be
U=u_{cl}\, \exp(i\xi)\, u_{cl},
\quad \xi=\sum_1 ^{N^2_f-1} {\xi^a \lambda^a \over F}\ .
\en
The terms linear in $\xi$ can then be made to vanish by imposing 
that the equation of motion be satisfied by the classical configuration,
\be\lbl{eom}
\left(\nabla _\mu u^\mu-{i\over2}\hat\chi_-
-{iC\over F^2}[\qp,\qm]\right)_{u=u_{cl}}=0,\quad
\hat\chi_-=\chi_- -{1\over N_f}\trace{\chi_-}\ .
\en
Next, from the terms quadratic in $\xi$ one defines the pion propagator
$G^{ab}_\Delta$ (we use the same notation as BCE ). It satisfies
the following equation (in euclidean space)
\be
(-d_\mu d_\mu +\sigma) G_\Delta(x,y)=\delta(x-y)
\en
where
\be
d_\mu=\partial_\mu +\gamma_\mu,\quad 
(\gamma_\mu)^{ab}=-{1\over2}\langle \Gamma_\mu[\lambda^a,\lambda^b]\rangle\ ,
\en
and
\bea\lbl{sigma}
&&\sigma^{ab}={1\over8}\langle[u_\mu,\lambda^a][u_\mu,\lambda^b]\rangle
+{1\over8}\langle\{\lambda^a,\lambda^b\}\chi_+\rangle\nonumber\\
&&\quad\quad
-{C\over4F^2}\langle [Q_+,\lambda^a][Q_+,\lambda^b]
                    -[Q_-,\lambda^a][Q_-,\lambda^b]\rangle\ .
\ena
Concerning the photon field, its classical part  is included
in the external field $v_\mu$ (which is not constrained by an
equation of motion) and the quantum part is designated by $A_\mu$. 
If we restrict ourselves to terms quadratic in the electric charge
we will only need the free propagator which, for
an arbitrary gauge parameter $\lambda$ reads
\be\lbl{photprop}
G^{\mu\nu}(x)=\delta^{\mu\nu} G_0(x)+(1-\lambda)
\int dz\,\partial^\mu\partial^\nu G_0(z)G_0(x-z)
\en
$G_0$ being the massless scalar free-field propagator. The remaining terms
quadratic in the fluctuating fields yield the two $O(p^2)$ vertices
(all vertices below are displayed in euclidian space)
\bea\lbl{p2ver}
&&{\cal L}^{(2)}_{\xi A}=-{F\over 2}\left(
(A_\mu\, d_\mu\xi^a)\trace{\lambda^a \qm}+
(A_\mu      \xi^a){i\over 2}\trace{\lambda^a [u_\mu,\qp]}\right)
\nonumber\\
&&{\cal L}^{(2)}_{A}={F^2\over 2} A_\mu\trace{\um\qm}
\nonumber\\
&&{\cal L}^{(2)}_{A A}=-{F^2\over4}A_\mu A_\mu \trace{\qm^2}
\ena

\subsection{$O(p^4)$ vertices}

Let us now turn to the vertices generated from the $O(p^4)$ Wess-Zumino-Witten
action. In the case of two flavours, to which we will restrict ourselves,
the action is local and
Kaiser\cite{kaiser} has recently derived a remarkably simple expression 
for the Lagrangian,
\be\lbl{wzwlag}
{\cal L}^W_{p^4}=\ceps\left(
\langle U^\dagger r_\mu U l_\nu-r_\mu l_\nu +
i\Sigma_\mu(U^\dagger r_\nu U +l_\nu)\rangle\langle v_{\alpha\beta}\rangle
+{2\over3}\langle \Sigma_\mu  \Sigma_\nu \Sigma_\alpha\rangle\langle 
v_\beta\rangle \right)
\en
with
\be
\kappa =-{N_c\over32\pi^2}\ 
\en
and 
\be
\Sigma_\mu=U^\dagger\partial_\mu U,\quad
l_\mu=v_\mu-a_\mu,\ r_\mu=v_\mu+a_\mu\ ,\ 
\epsilon^{0123}=-1\ .
\en
Performing the fluctuations we first obtain the vertices which are linear
in the quantum fields (calling $\hat\epsilon$
the $\epsilon$ tensor rotated to euclidian space),
\bea\lbl{linver}
&&{\cal L}^W_\xi={1\over4 F}\deps(\xi^a)\trace{\lambda^a (2iu_\mu u_\nu
+\fpmn)}\trace{\fpab}
\nonumber\\
&&{\cal L}^W_A={2\over3}\deps(A_\mu)
\trace{\Sigma_\nu\Sigma_\alpha\Sigma_\beta}
\trace{Q}\ .
\ena
Then, we obtain the three vertices which are quadratic,
\bea\lbl{p4ver}
&{\cal L}^W_{\xi\xi}=& {i\over4 F^2}\deps(d_\mu\xi^a \xi^b)
\trace{\com{\lambda^a,\lambda^b} u_\nu}\trace{\fpab}
\nonumber\\
&{\cal L}^W_{\xi A}=&{1\over2 F}\deps\Big\{-(A_\mu\xi^a)
\trace{\lambda^a (Q_{+\nu}+i\com{\qm,u_\nu})}\trace{\fpab}+
\nonumber\\
&&(\partial_\mu A_\nu \xi^a )\left[
\trace{\lambda^a\qp}\trace{\fpab}+
\trace{\lambda^a (\fpab+2iu_\alpha u_\beta)}\trace{\qp}
\right]\Big\}
\nonumber\\
&{\cal L}^W_{A A}=& 2\deps (A_\nu \partial_\alpha A_\beta)\left(
\trace{u_\mu \qp}\trace{\qp} -2\trace{a_\mu (Q_L+Q_R)}\trace{\qp}
\right)
\ena
One observes that the vertices  with either one or 
two $\xi's$ involve only canonical chiral building blocks. 
This implies that the contribution from graph (a) 
of fig.1 will obey 
ordinary chiral Ward identities\cite{wyler,bramon,issler}. This
property remains true for graphs (c) and (d). In contrast, the vertices 
which contain only $A_\mu$  generate non invariant pieces.  
It turns out, as we will see below,  
that these non invariant pieces do not affect the  
generating functional at order $e^2 p^4$ (except, however,
for the Coulomb term). The Adler-Bardeen non-renormalization
theorem\cite{bardeen1} is therefore satisfied in the 
effective theory at this order.  We have not investigated how the Adler-Bardeen
discussion is compatible with the effective theory at higher order in $e^2$.

\subsection{One-loop 1PI diagrams}

Let us now consider each diagram if fig.1 in turn. 

$\bullet$ diagram (a):

This diagram was considered in refs.
\cite{wyler,bramon,issler,aa} and the 
divergent part was extracted for generic $N_f$ but traceless external fields
and in 
ref.\cite{bgt} for the particular case of two flavours and 
non-traceless external vector fields. We consider the latter situation here
and want to extend the calculation to include the terms proportional to
$e^2 C$. The contribution from this diagram to the generating functional
in euclidean space is straightforward to obtain,
\be
-Z^{(a)}={i\over4 F^2}\deps\int dx\, d^x_\mu G^{ab}_\Delta(x,y=x) 
\trace{ [\lambda^a,\lambda^b]  u_\nu}\trace{\fpab},\quad
\en
The divergent part of the propagator can be obtained by heat-kernel methods
(see ref.\cite{bce2} and references therein) in dimensional regularization
\be
[d^x_\mu G_\Delta(x,y=x)]_{div}=(-2){1\over16\pi^2(d-4)}d^x_\mu a_1(x,y=x) 
\en
In the coincidence limit, the derivative of the heat-kernel coefficient 
$a_1$ is easily worked out to be
\be
d_\mu a_1=-{1\over6}(\partial_\lambda \gamma_{\lambda\mu}+[\gamma_\lambda,
\gamma_{\lambda\mu}])-{1\over2}(\partial_\mu\sigma +[\gamma_\mu,\sigma])\ .
\en 
The terms containing $\sigma$ will not contribute to the divergence because
they are symmetric in the $a,b$ indices. Performing the sum over $a,b$ and
returning to Minkowski space one finds that the divergence can be written
in terms of a Lagrangian,
\be
{\cal L}^{(a)}_{div}={1\over16\pi^2(d-4)}{-2\kappa\over 3F^2}
i\epsilon^{\mu\nu\alpha\beta}\langle \nabla^\lambda \Gamma_{\lambda\mu}
u_\nu\rangle \langle f_{+\alpha\beta}\rangle
\en
in agreement with the previous results\cite{wyler,bramon,issler,aa}. 
At this point, one notes that no
term proportional to $e^2 C$ appears. Such terms appear only at a later stage,
upon using the equation of motion. Therefore, they depend on the choice made
for the independent chiral Lagrangian terms in ${\cal L}^W_{p^6}$. 
We will use the basis of 
ref.\cite{bgt}. For convenience, we reproduce below the list of 
terms from this basis which are relevant for the divergence:
\be\lbl{bijinv}
\begin{array}{ll}
 o^W_6\equiv \eeps \trace{\fpmn} \trace{\cm\ua\ub}  \quad
& o^W_7\equiv i \eeps \trace{\fpmn} \trace{\fpab\cm}  
\\ \noalign{\vskip0.2truecm}
 o^W_8\equiv i \eeps \trace{\fpmn} \trace{\fpab}\trace{\cm}\quad   
& o^W_9\equiv i \eeps \trace{\fpgm} \trace{\hgn\ua\ub}  
\\ \noalign{\vskip0.2truecm}
 o^W_{10}\equiv \displaystyle i\eeps \trace{\fpgm} \trace{\fmgn\ua\ub}\quad  
& o^W_{11}\equiv \eeps \trace{\fpmn} \trace{\fpga\hgb}   
\\ \noalign{\vskip0.2truecm}
 o^W_{12}\equiv  \eeps \trace{\fpmn} \trace{\fpga\fmgb}\quad  
& o^W_{13}\equiv \eeps \trace{\nabla_\gamma \fpgm} \trace{\fpna\ub}\ .
\end{array}
\en
Let us also introduce two
additional chiral Lagrangian terms which appear if one does not make 
use of the equations of motion
\bea
&&o^W_{E1}\equiv i\eeps \trace{\fpmn} \trace{\nabla^\gamma u_\gamma
u_\alpha u_\beta}\nonumber\\
&&o^W_{E2}\equiv  \eeps \trace{\fpmn} \trace{\nabla^\gamma u_\gamma
\fpab}\ .
\ena

Not using the equation of motion, one finds that the divergence
from graph (a) has the following expansion
\be\lbl{divbgt0}
{\cal L}^{(a)}_{div}={-1\over16\pi^2(d-4)}{\kappa\over 6F^2}
\left( o^W_{E1} -o^W_{E2} -o^W_9+3o^W_{10}-2o^W_{11}+2o^W_{13}\right)\ .
\en
Using the equation of motion  eq.\rf{eom}, one obtains, firstly, the part which
does not depend on $e^2$,
\be\lbl{divbgt}
{\cal L}^{(a)}_{div1}={-1\over16\pi^2(d-4)}{\kappa\over 6F^2}
\left(
-{1\over2}o^W_6 -{1\over2}{o^W_7}+{1\over4}{o^W_8}
                         -o^W_9+3o^W_{10}-2o^W_{11}+2o^W_{13}\right)\ .
\en
This reproduces the result of ref.\cite{bgt} but one must keep in mind
that this does not represent the complete divergence because of the 
presence of 1PR contributions to be discussed below.  
Secondly, a part  proportional to $e^2 C$ appears, which can be expressed over 
the appropriate invariants $\ow{i}$ introduced above,
\be\lbl{divC}
{\cal L}^{(a)}_{div2}={-1\over16\pi^2(d-4)}{\kappa\over 6}
{C\over F^4} (-\ow{2} +\ow{5})\ .
\en

$\bullet$ diagram (b):

This diagram is easily seen to vanish since the vertex function is 
antisymmetric in two Lorentz indices which must be contracted with those of
the photon propagator which is symmetric.

$\bullet$ diagram (c):

In order to obtain the divergent part of this diagram one  uses the short
distance expansion of the pion propagator $G_\Delta(x,y)$ obtained from its
heat kernel representation,
\be
G_\Delta(x,y)=G_0(x-y) a_0(x,y)+...
\en
Only the first term will contribute to the divergence in the present case.
One also needs the divergent part in the product of two free propagators
\be
[G_0(x-y)^2]_{div}= {-2\over 16\pi^2 (d-4)}\delta(x-y)
\en
and the derivative of the preceding expression. Using the vertices from
eqs.\rf{p2ver}\rf{p4ver} a small calculation gives
the divergence in the following form in the Feynman gauge, $\lambda=1$
\bea\lbl{divc}
&{\cal L}^{(c)}_{div}=&{1\over 16\pi^2 (d-4)}{1\over2}\ceps\Big\{
\trace{\fpab}\trace{\,\left(\nabla_\mu\qm-i[\um,\qp]\right)  \times
\nonumber\\
&&\quad (\nabla_\nu\qp-{i\over2}[\un,\qm])}
+\trace{(\fpab+2i\ua\ub)\{\qp,Y_{\mu\nu}\} }\Big\}
\ena
where
\be
Y_{\mu\nu}=-{1\over2}[\Gamma_{\mu\nu},\qm]-{i\over4}[\fmmn,\qp]
-{i\over2}[\um,\nabla_\nu\qp]\ .
\en
This can be expressed in terms of our set of invariants
\be\lbl{divc1}
{\cal L}^{(c)}_{div}={-1\over 16\pi^2 (d-4)}{\kappa\over 8}
\left(4\ow{2}-\ow{3}-2\ow{4}+2\ow{7}-2\ow{8}\right)\ .
\en
$\bullet${Diagram (d)}

In this case, the divergence arises from the product of three free propagators
having two or more derivatives acting on them. The formula which is
needed here is
\be
[G_0(z-x)G_0(z-y)\partial_\mu\partial_\nu G_0(x-y)]_{div}=
{1\over2}{1\over16\pi^2(d-4)}\delta_{\mu\nu}\delta(x-y)\delta(x-z)\ .
\en
It is then not difficult to find the divergence from this diagram,
in the Feynman gauge
\be\lbl{divd}
{\cal L}^{(d)}_{div}={1\over 16\pi^2 (d-4)}{i\over4}
\ceps
\trace{\fpab}\trace{(\nabla_\mu\qm-{i\over2}[\um,\qp])[\un,\qm]}\ ,
\en
which, in terms of the basis of invariants read
\be
{\cal L}^{(d)}_{div}={-1\over 16\pi^2 (d-4)}{\kappa\over8}
\left( -2\ow{2} +2\ow{8}\right)
\en
The last diagram (e) is found to be convergent. In any case, it 
is quartic in the electric charge, which is beyond the
accuracy to which we restrict ourselves here. 

To summarize, the divergence
from the 1PI diagrams is contained in eqs.\rf{divC},\rf{divc} 
and \rf{divd}. An alternative method for determining this part of
the divergence 
is to express the one-loop generating functional as the trace-log
of an operator. We have performed the computation in this way also and
recovered the results of the diagrammatic approach. The diagrammatic 
method makes it somewhat easier to make use of gauges other than the 
Feynman gauge. It is not difficult, from the expression of the
photon propagator in an arbitrary gauge, eq.\rf{photprop} 
to determine the additional divergent part for $\lambda\ne1$
\be
{\cal L}^{(c+d)}_{div,\lambda}=(1-\lambda){-1\over 16\pi^2 (d-4)}
{\kappa\over16}\left( -4\ow{2}-\ow{5}+2\ow{7}+6\ow{8}\right)
\en
In the following, we will restrict ourselves to  $\lambda=1$ .

\subsection{One-loop 1PR diagrams}

Let us now consider the contributions to the $O(p^6)$ and $O(e^2p^4)$
divergence generated by one-particle reducible one-loop diagrams (see
fig.2). As is explained by BCE\cite{bce2} such contributions arise because
the divergences of the generating functional at one-loop  
are canceled by the contributions from the chiral $O(p^4)$ action up
to terms proportional to the equation of motion. If one uses the 
$SU(2)$ chiral Lagrangian of ref.\cite{gl84} extended to include dynamical
photons by Knecht and Urech (KU)\cite{ku}, one finds the following results
for these EOM terms,
\be\lbl{seom}
S^{EOM}\equiv S^{div}_{p^4}+S^{KU}_{p^4}=
{-1\over16\pi^2(d-4)}\int d^4x \Big(
{i\over2}\trace{ \chi^- X_2}-{iF^2\over8}\trace{\com{\qp,\qm} X_2} 
\Big)
\en
with
\be
X_2=\nabla^\mu u_\mu -{i\over2}\hat \chi_- -{iC\over F^2}\com{\qp,\qm}\ .
\en 
The one $\xi$ vertex resulting from the Lagrangian of eq.\rf{seom} reads
\be\lbl{onexi}
{\cal L}^{EOM}_{\xi}=
{-1\over16\pi^2(d-4)} (-d^2+\sigma)^{ab}\xi^b \left(
{i\over2F}\trace{\lambda^a\hat\chi_-} -{iF\over8}
\trace{\lambda^a\com{\qp,\qm}}
\right)\ .
\en
Together with the one $\xi$ vertex from ${\cal L}^W_{p^4}$, eq.\rf{onexi}, 
this produces the following local divergent terms of order $p^6$ and $e^2p^4$
from the diagrams of fig.2
\be
{\cal L}^{(f+g)}_{div}=
{-1\over16\pi^2(d-4)}\left[{\kappa\over2F^2}\left(-o^W_6 +{1\over2}o^W_7
-{1\over4}o^W_8\right)
+{\kappa\over8}\left(-\ow{2}-{1\over2}\ow{5}\right)\right]
\en

Finally, the UV divergences of the Green's functions at order
$e^2p^4$ can be absorbed into the bare coupling constants $k_i^W$. We will
adopt the same convention as ref.\cite{bgt} for the relation between
bare and renormalized couplings (which differs from the normalization
adopted in the ordinary $p^6$ sector\cite{bce2}),
\be\lbl{renorm}
k^W_{i,bare}=(c\mu)^{d-4}\left({1\over 16\pi^2(d-4)} \beta_i
+k^W_{i,r}(\mu)\right)\ ,
\en
with $\log c=-(\log{4\pi}+\Gamma'(1)+1)$. 
Collecting the various pieces in the divergences we find that the coefficients
$\beta_i$ have the following values in the Feynman gauge
\be
\beta_1=\beta_6=\beta_8 =0
\en
and
\be
\begin{array}{lll}
  \beta_2=\left({1\over8}-{1\over6}{C\over F^4}\right)\kappa,\ 
& \beta_3=-{1\over8} \kappa,\ 
& \beta_4=-{1\over4}\kappa,  \\ \noalign{\vskip0.3truecm}
  \beta_5=\left(-{1\over16}+{1\over6}{C\over F^4}\right)\kappa,\ 
& \beta_7= {1\over4}\kappa\ . 
\end{array}
\en
We also quote the values of the analogous parameters $\eta_i$
associated with the $O(p^6)$ couplings $c_i^W$ taking into account
the 1PI as well as the 1PR contributions,
\be
\begin{array}{lll}
 \eta_6=-{7\over12}\const,\ 
&\eta_7={1\over6}\const,\ 
&\eta_8=-{1\over12}\const,\  \\ \noalign{\vskip0.3truecm} 
 \eta_9=-{1\over6}\const,\ 
&\eta_{10}={1\over2}\const,\ 
&\eta_{11}=-{1\over3}\const,\ \\ \noalign{\vskip0.3truecm} 
\eta_{12}=0,
&\eta_{13}={1\over3}\const\ ,
\end{array}
\en
and $\eta_i=0,\ i=1,\ldots,5$ .
\subsection{Application to $\gamma\pi\to\pi\pi$}

These results can be applied to the computation of radiative corrections
at one-loop to various processes in the anomaly sector. Taking into account
the tree level contributions from ${\cal L}^W_{e^2p^4}$ enables one
to express the result in a finite and scale independent way in terms of
a minimal number of coupling constants. For instance, let us consider
the process $\gamma\pi^+\to\pi^+\pi^0$ for  which the one-loop radiative
corrections have already been calculated\cite{akt}. We follow the notation
of this reference and denote the amplitude for this process by $F^{3\pi}$,
\be
\trace{0\vert J^{em}_\mu(0)\vert \pi^+\pi^-\pi^0}=i\eeps
p_{0\nu}p_{+\alpha}p_{-\beta}\,F^{3\pi}(s_+,s_-,s_0)\ .
\en
The current algebra result for this amplitude is denoted by 
$F^{3\pi}_0$ and reads
\be
F^{3\pi}_0={-8e\kappa\over3 F^3_\pi}
\en
The tree-level contribution  from the Lagrangian
${\cal L}^W_{e^2p^4}$ to $F^{3\pi}$ is easily evaluated to be,
\be
F^{3\pi}_{tree}={-32 e^3\over 3 F^3}\left(
-2k^W_{2,r}+2k^W_{3,r}+k^W_{4,r} \right)\ .
\en
Using the results above one finds that this part has the following scale
dependence,
\be
\mu{d\over d\mu}F^{3\pi}_{tree}={e^2\over 16\pi^2} F^{3\pi}_0
\left({-4C\over3F^4} +3\right)\ .
\en
This scale dependence should cancel
that arising from the one-loop contributions to the amplitude. According
to ref.\cite{akt}, however,
\be
\mu{d\over d\mu}F^{3\pi}_{loop}={-e^2\over 16\pi^2} F^{3\pi}_0
\left({-8C\over F^4} +3\right)\ .
\en
Indeed we find the cancellation occuring for the $e^2$ term, but not for
the $e^2 C$ term.

\section{Chiral and electromagnetic corrections to $\pi^0\to2\gamma$}

We begin by setting up 
a complete list of the corrections to the current
algebra result for the $\pi^0$ decay amplitude  
including both quark mass and electromagnetic corrections. 
might, a priori, expect to be of comparable  magnitude. 
An investigation with a similar scope
was undertaken some time ago by Kitazawa\cite{kitazawa}.
Here, we will be  using the approach of ChPT. In this framework, all
corrections are expressed in terms of a minimal set of coupling-constants: at
a given order, it is garanteed that no effect has been forgotten or double
counted. We expect couplings from several sectors of ChPT to be involved:
couplings from ${\cal L}_{p^4}$, from ${\cal L}^W_{p^6}$ 
(it is important here
that the minimal number of independent couplings of this type has now been
correctly determined\cite{bgt,FS2}) and, concerning EM corrections, 
couplings from ${\cal L}_{e^2p^2}$ and from  ${\cal L}^W_{e^2p^4}$ that we
have discussed above.
In a next step, we will
discuss resonance saturation estimates for all the combinations of couplings
which are involved. Such estimates cannot be expected to be very accurate
but they do provide reliable orders of magnitude and this discussion will
show that, in fact, a single term dominates which one can then determine by
making use of the ChPT expression for the $\eta$ decay amplitude.

The T-matrix element for the $\pi^0$ decay into two photons has the following
structure
\be
\trace{\gamma(p)\gamma(q)\vert {\cal T}\vert \pi^0}=
\eeps p_\mu e_\nu q_\alpha e'_\beta\, A_{\pi2\gamma}\  
\en
in terms of the momenta and the polarization vectors of the two photons.
Let us call the current algebra result for the amplitude $A_{CA}$. It has
the well-known expression\cite{adler,belljackiw}
\be\lbl{ca}
A_{CA}={\alpha \over \pi F_\pi}\ 
\en
where $F_\pi$ is the pion decay constant. Let us recall how this quantity is
determined from experiment. As shown by Marciano and Sirlin\cite{ms} $F_\pi$
can be related to the charged pion decays $\pi^+\to \mu^+\nu, \mu^+\nu\gamma$
by the formula,
\bea\lbl{msform}
&&\Gamma(\pi^+\to\mu^+\nu(\gamma))=F^2_\pi \,{G^2_F \vert V_{ud}\vert^2
\over4\pi} m^2_\mu m_{\pi^+} \left(1- z^2_{\pi\mu}\right)^2\times
\nonumber\\
&&\quad\quad\left (1+{2\alpha\over\pi}\log{m_Z\over m_\rho}\right)
\left(1+{\alpha\over\pi} F(z_{\pi\mu})\right)
\left(1-{\alpha\over\pi}\left(-{3\over4}\log{m^2_\pi\over m^2_\rho}+c_1\right)
\right)
\ena
with $z_{\pi\mu}=m_\mu/m_{\pi^+}$ and the function $F$ has the following
expression
\bea
&&F(x)=3\log{x}+{13-19 x^2\over8(1-x^2)}-{8-5 x^2\over2(1-x^2)^2}\,x^2\log{x}
\nonumber\\
&&\qquad -2\left( {1+x^2\over1-x^2}\log{x}+1\right)\log(1-x^2)
+2 {1+x^2\over1-x^2}\,\int_0^{1-x^2}{\log(1-t)\over t} dt\ .
\ena
Further radiative corrections proportional to the ratio $m^2_\mu/m^2_\rho$
have been dropped. 
In eq.\rf{msform} all the dependence upon the electric charge $e^2$ is 
displayed explicitly such that $F_\pi$ is defined at $e^2=0$. 
In this situation, the difference
between $F_{\pi^+}$ and $F_{\pi^0}$ is quadratic in $m_u-m_d$, it 
can be shown to be
of the order of $10^{-4}$ and can be ignored for our purposes. 
The only undetermined parameter in eq.\rf{msform} is $c_1$. 
One can think of this parameter as collecting a number of ChPT low-energy
coupling-constants\footnote{An explicit expression in terms of such 
coupling-constants was derived in ref.\cite{knecht2}. }.
In ref.\cite{ms} its variation with the scale was used to estimate that it must
lie in a range $c_1\simeq 0\pm 2.4$. A more sophisticated analysis using
resonance models for the various form factors involved 
was undertaken by Finkemeier\cite{fink} who obtains $c_1=-3.0\pm 0.8$. 
Chiral perturbation theory shows that, at one-loop,  
there are electromagnetic effects which are induced by the $\pi^+-\pi^0$
mass difference\cite{knecht2}. 
The corresponding chiral logarithms are  not accounted
for by the resonance models\footnote{
These models provide a leading large $N_c$ approximation as, for instance,
the resonances are taken as infinitely narrow. }  
and we feel that they could be added explicitly
into $c_1$. In this way, together with that appearing in eq.\rf{msform} the 
complete set of  chiral logarithms is included. Finally, we will use
\be\lbl{c1}
c_1=-3.0\pm0.8+{C\over4 F^4_\pi}\left(3+2\log{m^2_\pi\over m^2_\rho}
+\log{m^2_K\over m^2_\rho}\right)\ ,
\en
where $C$ is the coupling which appears at chiral order $e^2$\cite{urech}. For
the other quantities we use the values from the PDG\cite{pdg2000}, i.e.
\be
\begin{array}{c}
\Gamma(\pi^+\to\mu^+\nu(\gamma))=0.25281(5)\,10^{-16}\ {\rm GeV},
\quad V_{ud}=0.9735(8)\ \\
G_F=1.16637(1)\,10^{-5}\ {\rm GeV^{-2}}\\
\end{array}
\en
This results in the following value for $F_\pi$ 
\be\lbl{fpival}
F_\pi=92.16\pm 0.11\ {\rm MeV}
\en
which we will use in what follows\footnote{
The difference with the value quoted in the PDG\cite{pdg2000}, 
$F_\pi=92.4\pm 0.3$ MeV, comes partly from our
using the value of $c_1$ from ref.\cite{fink} and partly from using the values
of $G_F$ and $V_{ud}$ provided by the same PDG. }.

Let us now consider the $\pi^0$ decay process from the point of view of ChPT. 
For this particular process, it will turn out to be fruitful to use 
the enlarged framework of the $SU(3)$ chiral expansion. 
This expansion, of course, 
is less general than the $SU(2)$ one, as it relies on the 
additional assumption that the strange quark mass is sufficiently small.  
In the$SU(2)$ ChPT at $O(p^4)$ the amplitude reads
\be
A_{p^4}={\alpha \over\pi F}\quad [SU(2)]
\en 
where $F$ is the pion decay constant in the limit $m_u=m_d=0$. 
In the $SU(3)$ expansion at $O(p^4)$, the amplitude receives 
a correction due to the $\pi^0-\eta$ mixing and reads
\be
A_{p^4}={\alpha\over\pi F_0} \left( 1+{m_d-m_u\over4(m_s-m)}\right),\quad
m=\undemi(m_u+m_d)\quad [SU(3)]
\en
where $F_0$ is the pion decay constant in the limit $m_u=m_d=m_s=0$.
At this level of the chiral expansion, one can identify $F$ 
and $F_0$ with $F_\pi$ but, still, the two amplitudes differ. 
For the problem at hand,  the  $SU(3)$ expansion 
enables one to make use of input from
the $\eta$ decay amplitude and this will result in much improved
predictions. Therefore, we will use the $SU(3)$ chiral expansion
in the following. 
Let us now list the contributions at the next chiral order.

\noindent 1)One-loop meson diagrams which are one-particle 
irreducible\footnote{The separate
contributions of one-particle reducible and irreducible diagrams depends
on the representations of the matrix $U$ and the fluctuation matrix
in terms of pion fields. For this matter, 
we will follow the conventions of sec.3 }. An exact
cancellation occurs between the tadpole and the unitarity
contributions \cite{dhl,bbc}. This still holds when $O(e^2)$ contributions
are included in the masses.
 
\noindent 2)One-particle irreducible tree contributions from 
${\cal L}^W_{p^6}$. In $SU(3)$, there are only two terms from the list
of ref.\cite{bgt} which contribute, $C^W_7$ and $C^W_8$,
\be
{\cal L}^W_{p^6}=\eeps\left\{ ...+C^W_7\trace{\chi_-\fpmn\fpab}
+C^W_8\trace{\chi_-}\trace{\fpmn\fpab}+...\right\}
\en
It is convenient
to introduce two related dimensionless parameters
\be
T_1=-{256\pi^2\over 3}m^2_\pi C^W_7\quad 
T'_1=-{1024\pi^2\over3}m^2_\pi C^W_8
\en
(we use the same notation as ref.\cite{bm} the difference in the sign
reflects the different convention for the epsilon tensor). If one 
were to perform the $SU(2)$ chiral expansion, there would be 
four independent coupling constants
involved: $c^W_3$, $c^W_7$, $c^W_8$ and $c^W_{11}$.\\

\noindent 3)One-loop and tree contributions which are one-particle
reducible. These can be expressed in terms of wave-function renormalization
and mixing. Including electromagnetic contributions but neglecting 
terms which are quadratic in isospin breaking (i.e. $(m_u-m_d)^2$, $e^4$, 
$e^2 (m_u-m_d)$ ) wave function renormalization has the following form
\be \lbl{mix}
\left(\begin{array}{c}
\pi^3\\
\    \\
\pi^8\\ 
\end{array}\right)=
\left(\begin{array}{cc}
1                       & -\epsilon_1-e^2\delta_1\\
\                       &\                        \\
\epsilon_2+ e^2\delta_2 & 1                  \\
\end{array}\right)
\left(\begin{array}{c}
{F_0\over F_\pi}(1+e^2\delta_\pi)\pi^0\\
\                                     \\
{F_0\over F_\eta}(1+e^2\delta_\eta)\eta\\
\end{array}\right)\ .
\en
The expressions for $\epsilon_1$, $\epsilon_2$ 
were given in ref.\cite{gl85}, we
reproduce the formula for $\epsilon_2$ here which is relevant for our 
calculations
\bea
&&\epsilon_2={\sqrt3(m_d-m_u)\over4(m_s-m)}\Bigg\{ 1-{32(\mkd-\mpid)\over
\fpid}(3L_7+L_8)-3\mu_\pi+2\mu_K+\mu_\eta\nonumber\\
&&\qquad +{\mpid\over16\pi^2\fpid}\left(1-{\mpid\over\mkd-\mpid}
\log{\mkd\over\mpid} \right)\Bigg\}\ 
\ena
where
\be
\mu_P= {m^2_P\over32\pi^2 F^2_0} \log{m^2_P\over\mu^2},\ 
P=\pi,\ K,\ \eta  \ .
\en
The electromagnetic contributions which will be relevant for us
are contained in $\delta_2$ and $\delta_\pi$ in eq.\rf{mix} which have
the following expressions
\bea\lbl{deltas}
&&\delta_\pi=-\undemi C_1 +{C\over\fpiq }(4\tilde\nu_\pi+\tilde\nu_K)
\nonumber\\
&&\delta_2={\sqrt3\, m\over m_s-m}\left({3\over2} C_2-{2\over3}(K_9+K_{10})
-{2C\over\fpiq} \tilde\nu_K\right)
\ena
where
\bea\lbl{Cdef}
&&C_1={8\over3}(K_1+K_2)-2(2K_3-K_4)+{20\over9}(K_5+K_6)
\nonumber\\
&&C_2=\phantom{{8\over3}(K_1+K_2)}
-{2\over3}(2K_3-K_4)+{4\over9}(K_5+K_6)
\ena
and
\bea
&&\tilde\nu_P={1\over32\pi^2}\left(
\log{m^2_P\fpid+2e^2C\over\mu^2\fpid}
+ {m^2_P\fpid\over 2e^2C}\log \left(1+{2e^2C\over m^2_P\fpid}\right)
\right)
\nonumber\\
&&\phantom{\tilde\nu_P}
\simeq {1\over32\pi^2}\left(1+\log{m^2_P\over\mu^2}\right)\ .
\ena
The last equality holds if one uses physical values for the quark masses
and the electric charge. 
For completeness we also quote the expressions for the remaining two 
parameters $\delta_\eta$ and $\delta_1$ in the mixing matrix
\bea
&&\delta_\eta=-\undemi C_1 +C_2+{3C\over\fpiq}\tilde\nu_K
\nonumber\\
&&\delta_1={\sqrt3\over m_s-m}\left(\undemi(2m_s+m) C_2-{2\over3}m
(K_9+K_{10})-{2C\over\fpiq}m_s\tilde\nu_K\right)
\ena 
The electromagnetic contributions to $\pi-\eta$ mixing were previously
discussed in ref.\cite{neufeld}. Their influence in the 
$\eta\to3\pi$ decay rate was investigated in ref.\cite{kambor}.\\

A remark is in order here concerning the
QCD renormalization group invariance of the combination $\epsilon_2+
e^2\delta_2$. The analysis of ref.\cite{bmdash} shows that $K_9+K_{10}$
which appears in eq.\rf{deltas} for $\delta_2$ depends not only on the
chiral scale $\mu$ but also on the QCD renormalization scale $\mu_0$. However,
$\epsilon_2$ also is not RNG invariant because $m_u$ and $m_d$ have different 
charges. The scale dependence is proportional to $e^2 m/(m_s-m)$ and it
can be seen to cancel out exactly in the combination $\epsilon_2+
e^2\delta_2$.

\noindent 4)There are no one-loop diagrams
with one photon in the loop. The last possible contributions are 
tree level contributions from ${\cal L}^W_{e^2p^4}$. Going through the
list of independent terms eq.\rf{lwe2p4} one sees that none of them 
contributes to the $\pi^0$ decay amplitude. These terms correspond to the
$SU(2)$ expansion but this result is easily seen to hold for the
$SU(3)$ expansion as well. 

Putting all this together, we can write the $\pi^0\to2\gamma$ amplitude
including all the contributions which are of chiral order $p^6$ or 
$e^2p^4$, i.e. linear in the quark masses $m_u$, $m_d$ and
the electric charge $e^2$ 
\bea\lbl{Api}
&&A_{\pi2\gamma}=A_{CA}\Big\{ 1+
\nonumber\\
&&\qquad {\epsilon_2\over\sqrt3} +{e^2\delta_2\over\sqrt3} +
e^2\delta_\pi 
+\left(   1-{m_d-m_u\over m}{4r-5\over4(r-1)} \right)T_1
-{3(m_d-m_u)\over 4m}T'_1 \Big\}\ 
\ena
where $r$ is the quark mass ratio 
\be
r={2m_s\over m_u+m_d}
\en 
Altogether, there are five independent
contributions which generate the deviation from the current algebra result.
These contributions are of same chiral order but, in practice, they can
have rather different sizes. This may be seen by using resonance models
to estimate the orders of magnitude of the various chiral coupling constants
which are involved. How this can be performed in practice is described in
detail in the appendix. Using these results, one obtains the following
estimates for the various entries in eq.\rf{Api}
\bea\lbl{estim}
&&{\epsilon_2\over\sqrt3}\simeq0.6\,10^{-2},\quad 
{e^2\delta_2\over\sqrt3}\simeq-1.2\,10^{-6},\quad
e^2\delta_\pi\simeq -0.3\,10^{-2}
\nonumber\\
&&\phantom{{\epsilon_2\over\sqrt3}\simeq0.6}
\vert T_1\vert <0.16\,10^{-2},\quad 
\vert T'_1\vert\simeq 3.04\,10^{-2}\ .
\ena
The term proportional to $\delta_2$ is negligible and the one containing $T_1$
is small. The latter term is the only one which would survive 
in the isospin symmetric limit. The corrections are entirely dominated by
isospin violation induced either by $m_u-m_d$ or by $e^2$. 
The dominant contribution in eq.\rf{estim} comes from $T'_1$. The resonance
models do not determine the sign of this contribution. A more accurate
determination including the sign can be performed based on the $SU(3)$ ChPT
expression at order $p^6$ of the $\eta\to2\gamma$ decay width,
\be\lbl{eta}
A_{\eta2\gamma}={1\over\sqrt3}A_{CA}\left\{ {F_\pi\over F_\eta} +
{5-2r\over3} T_1 +(1-r)T'_1\right\}\ .
\en
The decay constant $F_\eta$ which appears here is not known from experiment  
but can be determined using $F_\pi$, $F_K$ which are known, together with
ChPT\cite{gl85}. Next, using the experimental result for the $\eta$ decay
width $\Gamma_{\eta2\gamma}=0.46\pm0.04$ KeV and the fact that $T_1$ 
is much smaller than $T'_1$ one can determine this parameter in terms
of the quark mass ratio $r$ 
(assuming  that the sign of the amplitude is the same as its
current algebra value),
\be
T'_1={1\over1-r}(0.93\pm0.07\pm0.14)\ .
\en
Here, the first error reflects the experimental error on the width and the
second error reflects the uncertainty coming from 
higher order chiral corrections
in the formula \rf{eta}. We have  assumed that such corrections do not
to exceed 15\%. One indication that higher order corrections are not large
is that the value of $T'_1$ obtained using eq.\rf{eta}, 
$T'_1=(-3.7\pm0.3\pm0.4)\,10^{-2}$ (with the quark mass ratio taken
to be $r=26$) agrees with the estimate obtained using resonance 
models (see \rf{estim}). Using this evaluation of $T'_1$ one can write
the pion decay amplitude beyond current algebra as follows
\be\lbl{Rdet}
A_{\pi2\gamma}=A_{CA}\left\{1+{m_d-m_u\over m_s-m}
(0.93\pm0.12) -0.34\,10^{-2} \pm 0.14\,10^{-2}\right\}\ .
\en
In this formula, we have collected first the terms proportional
to $m_d-m_u$ which involve the inverse of the quark mass ratio 
called $R$,
\be
R={m_s-m_u\over m_d-m_u}\ .
\en
The term which comes next is the electromagnetic contribution
and the error (last term) collects the uncertainties in the estimate
of this contribution and the uncertainty from $T_1$.
The quark mass ratio $R$ has been determined previously from several 
sources:
a)leading order ChPT fit to the pseudo-scalar octet meson 
masses\cite{weinbergR} (in particular the $K^0-K^+$ mass difference)
giving $R\simeq 43$, b)from $\rho-\omega$ 
mixing (e.g.\cite{urechmix}  for a recent re-evaluation) and 
c)from the $\eta\to3\pi$ decay in ChPT (see the
discussion by Leutwyler\cite{leutrev}). The latter method seems to yield a
slightly smaller value of $R$. Agreement of various determinations would
imply that the expansion in the strange quark mass is rapidly converging
and would eliminate the possibility that $m_u$ is equal to zero (which gives
a value of $R$ of the order of twenty). In this sense it would be interesting
if a precise measurement of the $\pi^0$ decay width could yield such a new
determination upon using the formula \rf{Rdet}. Unfortunately, the smallness
of the effect and the theoretical uncertainty only allow for the order of
magnitude of $R$ to be tested. Using the value $R=43$ the final
theoretical prediction for the $\pi^0\to2\gamma$ decay width is
\be
\Gamma_{\pi2\gamma}=8.06\pm0.02\pm0.06 \ {\rm eV}\ .
\en
The first error is associated with the determination of $F_\pi$ 
(eq. \rf{fpival} ) and the second error
collects the uncertainties in the evaluation of the chiral corrections.
The overall accuracy is of the order of 1\%.

\section{Conclusions}
In this paper, we have discussed the framework for including radiative  
corrections
in chiral perturbation theory in the sector of the anomaly. For this
purpose, we have first classified a minimal set of chiral lagrangian
terms at $O(e^2p^4)$. In the case of $N_f=2$ there are eight such independent
terms. Next, we have computed the divergence structure at this order.
Contributions not only from irreducible but also from reducible
one-loop graphs are present, which conforms to the general discussion of
the $O(p^6)$ type divergences in ref.\cite{bce2}. We have used a diagrammatic
method which allows one to make use, eventually,  
of gauges more general than the Feynman gauge.
The effective action at $O(e^2p^4)$ is non-anomalous, similar to the
results obtained previously for the other $O(p^6)$ components in the
anomaly sector\cite{wyler,bramon,issler}. 
We expect that this property will no longer hold at higher
order in $e^2$. 

As a practical application, we have verified the divergence
structure which was obtained from an explicit calculation of EM corrections
to the $\gamma\pi\to\pi\pi$ amplitude: we find only partial agreement. As an
other application, we have considered in some detail the $\pi^0$ decay
amplitude which is of interest in view of forthcoming high-precision
measurements at JLAB. In this case, the coupling-constants from
${\cal L}^W_{e^2p^4}$ turn out to make no contributions. 
Under the assumption that the $SU(3)$ chiral expansion may be used, the
number of coupling-constants involved remains tractable. 
In this case, the whole set
of corrections to the current algebra result consists of five independent
terms, accounting for both quark mass and electromagnetic corrections: 
this is the content of eq.\rf{Api}. 
Estimates based on resonance models of the various
chiral coupling-constants involved are reviewed and updated and a quantitative
result is presented. The corrections are dominated by isospin breaking, 
with the electromagnetic contributions being small but not negligible.  

{\large\bf Acknowledgements: }

One of us (BM) would like to thank the CTS in Bangalore for its hospitality.
This work is supported in part by the Indo-French collaboration contract
IFCPAR 2504-1, in part by the Department of Science and Technology,
Government of India under the project entitled ``Some aspects of
Low-energy Hadronic Physics''
and by the EEC-TMR contract ERBFMRXCT98-0169 .

\section{Appendix: Resonance models of chiral coupling constants}

The sizes of chiral coupling constants in ChPT can be understood, at
a semi-quantitative level, in terms of  properties of the light resonances.
This was investigated in great detail by Ecker et al.\cite{egpr} 
for the case of the couplings $L_i$ of the $O(p^4)$ chiral Lagrangian. 
In this approach one introduces first a Lagrangian which includes, 
in addition to chiral fields and sources,  resonances
with proper transformation properties under the chiral group. 
The dynamics can be treated at tree level and one can match the Green's 
functions computed in this way with those obtained from ChPT.
Let us  discuss the predictions
of such an approach for the set of chiral coupling constants relevant for 
the $\pi^0$ decay  amplitude. 

\subsection{Couplings $C_1$, $C_2$}

Let us consider first the  
Urech's couplings $K_i$. Quite generally, each of these couplings can
be expressed as a QCD N-point Green's function (with N=2,3,4) convoluted
with the photon propagator\cite{bmdash}. In some cases, informations
on the relevant Green's functions can be extracted from experiment.
Otherwise, one may use  models for these. 
Several papers have discussed estimates for some or combinations of
$K_i's$\cite{bu,bijnens,donoper,bmdash} 
but we could not find a result for the
specific combinations $C_1$, $C_2$ (see eq.\rf{Cdef} )
which are needed here. In this case, a straightforward approach 
to obtain the  resonance model estimate is the following.

One can easily identify the  combinations $C_1$, $C_2$ 
by considering the effective action 
for the neutral pseudo-scalar mesons $\pi^3$, $\pi^8$ 
taken as slowly varying fields. In ChPT at order $e^2p^2$ the leading
terms in an expansion in powers of the derivatives has the following form
\be\lbl{Schir}
{\cal S}_{eff}= \undemi e^2 
\int d^4x\,\big\{ K_{11} \pitl\pith +2K_{12}\pitl \pieh 
+K_{22}\piel\pieh +...\big\}
\en
with $ \nabla_\mu\pi^i=\partial_\mu \pi^i-a^i_\mu$ and
\bea\lbl{Kij}
&&K_{11}=C_1 -{2C\over\fpiq}(4\tilde\nu_\pi+\tilde\nu_K)
\nonumber\\
&&K_{12}=\sqrt3 ( C_2-{2C\over\fpiq}\tilde\nu_K )
\nonumber\\
&&K_{22}=C_1-2C_2 -{6C\over\fpiq}\tilde\nu_K\ .
\ena

We will next
generate the  effective action for the neutral pseudo-scalars starting from
the following Lagrangian containing  vector meson resonances\cite{eglpr},
\bea\lbl{vlag}
&& {\cal L}_V=-{1\over4}\trace{ V_{\mu\nu} V^{\mu\nu} -2 m^2_V V_\mu V^\mu}
-{1\over2\sqrt2}f_V\trace{V_{\mu\nu} f_+^{\mu\nu}}
\nonumber\\
&& +{g_1\over2\sqrt2}\,\eeps \trace{ \{\um, V_\nu\}\fpab }
   +{g_2\over2      }\,\eeps \trace{ \{\um, V_\nu\} V_{\alpha\beta }}\ .
\ena
with $V_{\mu\nu}=\nabla_\mu V_\nu -\nabla_\nu V_\mu$. 
We use a description (introduced in ref.\cite{eglpr}) 
in terms of vector fields which transform 
homogeneously under the the non-linear representation of the chiral group,
\be
V_\mu \to h V_\mu h^\dagger
\en
This representation is related by a simple field transformation \cite{eglpr} 
to the hidden gauge approach\cite{bando}. As shown in ref.\cite{eglpr} 
different representations will give the same answer provided matching to the
QCD asymptotic behaviour is imposed. 
There are three types 
of one-loop diagrams which have one photon in the loop, they are shown
in fig.2.
\begin{figure}[abt]
\epsfysize=9cm
\begin{center}
\epsffile{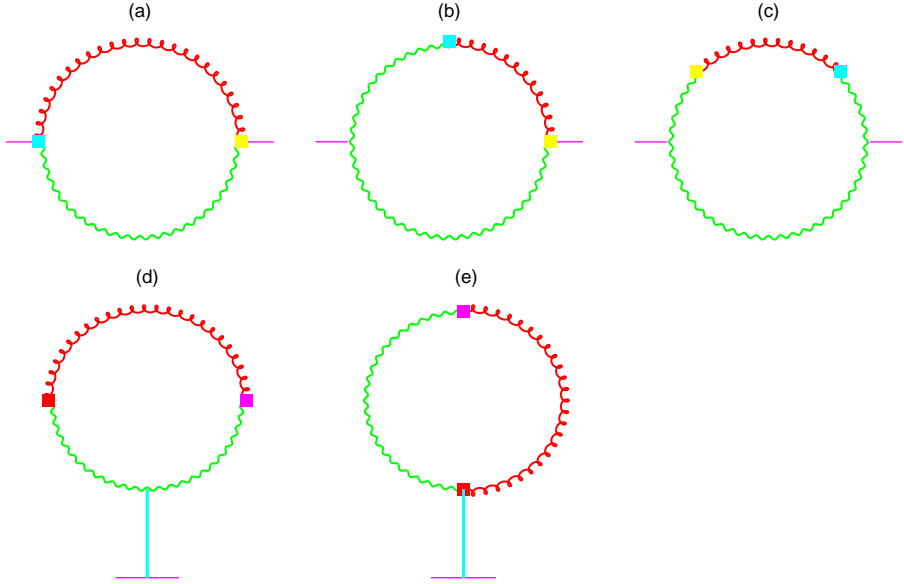}
\caption{ Graphs which contribute to order $e^2$ wave-function 
renormalization in resonance models. Graphs (a), (b), (c) are generated
by the vector meson Lagrangian eq.\rf{vlag}. A curly line denotes a
photon propagator and a wiggly line a vector meson propagator.
Graphs (d), (e) show 
possible contributions from scalar resonances which we have not included.}
\label{Fig. 3}
\end{center}
\end{figure}
Computing these diagrams 
gives the following result for the effective action
\be\lbl{Sres}
S_{eff}={4e^2\over F^2}\int d^4x\left\{ {10\over 9}\pitl\pitlnu 
+{4\over3\sqrt3}
\pitl\pielnu+{2\over3}\piel\pielnu \right\}I^{\mu\nu}+...
\en
with
\be
I^{\mu\nu}=i\int {d^np\over(2\pi)^n}{p^2g^{\mu\nu}-p^\mu p^\nu\over
p^2(p^2-m^2_V)^3 }\left(p^2\,(g_1+2g_2 f_V)-g_1 m^2_V\right)^2\ .
\en
In the present case, matching to the QCD asymptotic behaviour is equivalent
to requiring that the integral $I^{\mu\nu}$ converges (see \cite{bmdash}). 
This is satisfied provided the following relation holds
\be\lbl{grel}
g_1+2g_2 f_V=0\ .
\en
One can verify that the same relation ensures that the matrix element of the
vector current between a vector meson and a pion satisfies vector meson
dominance, i.e. has a simple pole form. Using eq.\rf{grel} one obtains
\be
I^{\mu\nu}=g^{\mu\nu}\, {3g^2_1 m^4_V \over128\pi^2} \ .
\en
We must next match the effective action obtained in ChPT eq.\rf{Schir}
and the one obtained from the resonance model eq.\rf{Sres}. Here, we face 
a problem: the chiral effective action has chiral logarithms which are
not present in the resonance model. This is due, in particular, to the
use of infinitely narrow resonances in the model. A plausible resolution
is to assume that the resonance model represents the part of the chiral action
without the chiral logarithms, i.e. the part which contains the combination
of couplings $C_1$, $C_2$ with the value of the scale set at $\mu=m_V$. 
One obtains, then
\be
C_1={5\,\tilde g^2_1\over 24\pi^2},\quad
C_2={\tilde g^2_1\over 24\pi^2},\quad
\tilde g_1={g_1 m_V\over F_\pi}\ .
\en
The dimensionless coupling constant $\tilde g_1$ can be determined from the 
$\omega\to\pi\gamma$ decay width $\Gamma=0.72\pm 0.05$ MeV,
\be
\Gamma=\alpha\, {\tilde g^2_1 m_V\over 6} 
\left(1-{m^2_\pi\over m^2_V}\right)^3,\quad
\tilde g_1 =0.91\pm 0.03\ .
\en
In ref.\cite{bmdash} it was shown that the Urech's couplings $K_1$,..$K_6$
can be expressed as convolutions involving QCD four-point functions. The
estimate discussed above corresponds to a vector meson pole model 
for the relevant four-point functions. Other possible poles, for instance
involving scalar mesons (see fig.3 ), are ignored. In eq.\rf{deltas} beside
$C_1$ and $C_2$ one also needs the combination $K_9+K_{10}$. We will use for
this the resonance estimate given in ref.\cite{bmdash}. 

\subsection{Couplings $C^W_7$, $C^W_8$ }

Let us now consider the two coupling-constants $C^W_7$, $C^W_8$ from
${\cal L}^W_{p^6}$. For the sake of using a unified approach we review
the estimates obtained using the same kind of resonance models 
as discussed above. The relevant resonances now are the pseudo-scalar ones,
we consider an octet of these ($\pi(1300)$ family) and two singlets
(corresponding to $\eta'(980)$ and $\eta(1295)$ ). Using the same notation
as ref.\cite{egpr} the couplings to the pseudo-scalar sources
are contained in
\be
{\cal L}_P= id_m\trace{P \chi_-}+i\tilde d_m\eta_1 \trace{\chi-}
+i\tilde d'_m \eta'_1 \trace{\chi-}
\en
In addition, we have to consider the couplings to two photons
\be
{\cal L}_{P2\gamma}=\eeps \{ g_{\pi'} \trace{P \fpmn\fpab}
+g_{\eta_1}\eta_1 \trace{\fpmn\fpab} +g_{\eta'_1}\eta'_1\trace{\fpmn\fpab}\}\ .
\en
Next, one has to integrate out the resonances to the order needed to generate
terms belonging to ${\cal L}^W_{p^6}$. 
This is easily performed by making the field redefinitions
\be
P\to P+{i d_m\over M^2_P}\left(\chi_- - {1\over N_f}\trace{\chi_-}\right)\quad
\eta_1\to\eta_1 +{i\tilde d_m\over M^2_{\eta_1}}\trace{\chi_-}
\en
and similarly for $\eta'_1$. In this way, the following estimates of the
couplings $C^W_{7,8}$ in terms of resonance parameters emerge,
\be\lbl{L78sat}
C^W_7={g_{\pi'}d_m\over M^2_P}\quad
C^W_8={g_{\eta_1}\tilde d_m\over M^2_{\eta_1}}
+{g_{\eta'_1}\tilde d'_m\over M^2_{\eta'_1}}
-{g_{\pi'}d_m\over N_f M^2_P}\ .
\en

It is possible to express $C^W_{7,8}$ in a more general way.
In the case of  $C^W_{7}$, for instance, consider the matrix element
of the pseudo-scalar current between two photon states,
\be
\trace{\gamma(p)\vert j^3_P(0)\vert \gamma(q)}=
\eeps p_\mu e_\nu q_\alpha e'_\beta\, F_P(t),\quad t=(p-q)^2
\en
with $j^3_P=i\bar\psi\gamma^5 {\lambda^3\over2}\psi$. Then, it is possible
to express  $C^W_{7}$ as an unsubtracted dispersive 
representation\cite{kitazawa},
\be
B_0 C^W_{7}=-{3\over64\pi}\int_{9m^2_\pi}^\infty {dt'\, Im F_P(t')
\over t'} ,
\en
an analogous representation also holds for  $C^W_{8}$ . These representations
are convergent because asymptotically, the perturbative QCD contribution
reads,
\be
Im F^{QCD}_P(t)=-{2\over3}{\alpha m\over t}\log{ \sqrt{t}+\sqrt{t-4m^2}
\over 2m}(1+O(\alpha_s) )
\en
and convergence is even faster in the chiral limit $m=0$. Therefore, the 
approximation to retain the lightest resonance contribution in the integrand 
which, in the narrow width approximation reproduces the results of the
resonance model discussed above, should be quite reasonable. 
In the expression \rf{L78sat} for $C^W_{8}$ 
the last two terms cancel exactly in the leading large $N_c$ limit and
we expect the first contribution, from the $\eta'$ to be largely dominant. 
The parameters $\tilde d_m$ and $M_{\eta_1}$ have been estimated in 
ref.\cite{egpr}
\be
\vert \tilde d_m\vert\simeq 20\ {\rm MeV}\quad
M_{\eta_1}\simeq 804 \ {\rm MeV}\ .
\en
There remains to evaluate the parameter, $d_m$, which was not done in 
ref.\cite{egpr}. For this purpose, we may impose on the resonance model
to match the QCD asymptotic behaviour for the Green's function
\be 
\Pi_{SP}(p^2)=i\int d^4x\, \exp(ipx)
\trace {0\vert T( j^3_S(x) j^3_S(0)-j^3_P(x)j^3_P(0))\vert 0}
\en
which is easily shown to go like $1/(p^2)^2$ in the chiral limit. This
imposes the following Weinberg-type relation between $d_m$ and its
scalar counterpart $c_m$
\be
d^2_m + {\fpid\over 8}= c^2_m \ .
\en
In addition to that, we may use the relation between $d_m$, $c_m$ and
the low-energy coupling-constant $L_8$ (see \cite{egpr}) in the same
resonance saturation model,
\be
L_8={c^2_m\over 2M^2_S}-{d^2_m\over 2M^2_P}\ .
\en
Taking $L_8\simeq 0.9\,10^{-3}$ gives,
\be\lbl{cmval}
\vert c_m\vert\simeq 51 \ {\rm MeV},\quad
\vert d_m\vert\simeq 39 \ {\rm MeV}\ .
\en
Next, the parameters  $g_{\pi'}$, $g_{\eta_1}$, can be deduced from the 
two-photons decay rates of the $\pi(1300)$ and the $\eta'$,
\be
\Gamma_{\pi'2\gamma}= \alpha^2 g^2_{\pi'} {128\pi\over9} m^2_{\pi'},\quad
\Gamma_{\eta'2\gamma}= \alpha^2 g^2_{\eta_1} {1024\pi\over9} m^2_{\eta'}\ .
\en
The experimental values for these  decay widths are\cite{pipexp,pdg2000}
\be
\Gamma_{\pi'2\gamma}< 0.1\ {\rm KeV},\quad 
\Gamma_{\eta'2\gamma}=   4.29\pm 0.15
\ {\rm KeV} \ .
\en
This determines all the necessary ingredients in the expressions \rf{L78sat}
for $C^W_{7,8}$ and 
one deduces the following result for the dimensionless
quantities $T_1$ and $T'_1$ proportional to $C^W_{7}$ and $C^W_{8}$, 
\be
\vert T_1\vert< 0.16\,10^{-2},\quad
\vert T'_1\vert\simeq 3.04 \,10^{-2}\ .
\en

\end{document}